\documentclass[conference, a4paper]{IEEEtran}

\IEEEoverridecommandlockouts
\usepackage[utf8]{inputenc}
\usepackage{makeidx}  
\makeindex
\usepackage[T1]{fontenc}
\usepackage[pdftex]{graphicx}
\usepackage{epstopdf}
\usepackage{cite}
\usepackage{url}
\usepackage[font=footnotesize]{caption}
\usepackage{xcolor}
\usepackage{amsmath}
\usepackage{amssymb}
\usepackage{multirow}
\usepackage{eso-pic}
\usepackage{balance}
\usepackage{listings}
\usepackage{xspace}
\usepackage{subcaption}

\usepackage[]{algorithm2e}
\DontPrintSemicolon
\SetAlCapSkip{1em}
\SetKwInOut{Input}{input}
\SetKwInOut{Output}{output}
\SetKwFunction{Print}{print}
\SetKwFunction{AllPaths}{allPaths}
\SetKwFunction{End}{end}

  \newlength{\figsize}
  \newlength{\subfigwidth}
  \newlength{\subfiglabelwidth}
\newcommand\fig[1]{Figure~\ref{fig:#1}}

\newcommand\sect[1]{Section~\ref{sec:#1}}

\newboolean{makevspace}
\newcommand{\cvspace}[1]{%
    \ifthenelse
        {\boolean{makevspace}}
        {\vspace{#1}}
        {}%
    }
  \newcommand{\rem}[1]{\textcolor{red}{\sout{#1}}}
     \renewcommand{\rem}[1]{}

\begin{document}
  
  \title{P4-MACsec: Dynamic Topology Monitoring and Data Layer Protection with MACsec in P4-SDN}
  
  \author{
      \IEEEauthorblockN{
        Frederik~Hauser,~
        Mark~Schmidt,~
        Marco~Häberle,~
        and Michael~Menth
    \thanks{This work was supported by the bwNET100G+ project which is funded by the Ministry of Science, Research and the Arts Baden-Württemberg (MWK).
        The authors alone are responsible for the content of this paper.}
      }
      \IEEEauthorblockA{
          Chair~of~Communication~Networks,
          University~of~Tuebingen,
          Tuebingen,
          Germany\\
    Email:
          \{%
        frederik.hauser,%
        mark-thomas.schmidt,%
        marco.haeberle,%
        menth\}@uni-tuebingen.de,
      }
  }
  
  \maketitle
  
  \pagenumbering{gobble}
  
  \begin{abstract}
We propose P4-MACsec to protect network links between P4 switches through automated deployment of MACsec, a widespread IEEE standard for securing Layer 2 infrastructures.
It is supported by switches and routers from major manufacturers and has only little performance limitations compared to VPN technologies such as IPsec.
P4-MACsec introduces a data plane implementation of MACsec including AES-GCM encryption and decryption directly on P4 switches.
P4-MACsec features a two-tier control plane structure where local controllers running on the P4 switches interact with a central controller.
We propose a novel secure link discovery mechanism that leverages protected LLDP frames and the two-tier control plane structure for secure and efficient management of a global link map.
Automated deployment of MACsec creates secure channel, generates keying material, and configures the P4 switches for each detected link between two P4 switches.
It detects link changes and performs rekeying to provide a secure, configuration-free operation of MACsec.
In this paper, we review the technological background of P4-MACsec and explain its architecture.
To demonstrate the feasibility of P4-MACsec, we implement it on the BMv2 P4 software switch and validate the prototype through experiments.
We evaluate its performance through experiments that focus on TCP throughput and round-trip time.
We publish the prototype and experiment setups on Github.
\end{abstract}
  
  \IEEEpeerreviewmaketitle
  
  \graphicspath{{figures/}}
  
  \section{Introduction}
\label{sec:introduction}

MACsec is a widespread IEEE standard that protects the Layer 2 with cryptographic integrity checks or symmetric encryption.
MACsec prevents man-in-the-middle attackers from inspecting, inserting or even modifying network packets that are transmitted between two network peers.
In contrast to VPN technologies such as IPsec, MACsec processing is implemented on forwarding chips without notable overhead in line rate performance \cite{cisco-macsec-innovation}.
Packets are protected in a point-to-point manner between MACsec peers so that control plane functions targeting higher layers can be still applied.
Although mechanisms for distributed key exchange exist, MACsec deployment is still time-consuming and complex.
It requires knowledge about the network topology, large efforts in switch configuration, and typically maintenance of a key server.
Currently, automated deployments using a network management system with legacy switches is not feasible.
Legacy network switches only support the Link Layer Discovery Protocol (LLDP) that lacks in detecting topology changes in a timely manner.
In addition, it is vulnerable to several attacks that may result in a incorrect topology view.
Also, current legacy network switches do not support an automated configuration of MACsec through a southbound protocol.
Although a MIB for manipulating MACsec configuration with SNMP exists \cite{macsec-mib}, only basic MACsec parameters can be modified so that distributed protocols and a key server would be still required.

Software-defined networking (SDN) splits the strong binding between data and control plane.
OpenFlow (OF) \cite{openflow-paper} is the most widespread standard for SDN.
It consists of SDN switches with a fixed-function data plane that are steered by a central SDN controller.
P4 \cite{p4-paper} is a novel domain-specific language that introduces programmability to the data plane of P4-capable packet forwarding devices such as ASICs, CPU-based targets, and FPGAs.
Data plane behaviour can be described in P4 programs that run on P4 switches so that network operators can constantly program the packet processing on deployed switches.
The P4 Runtime \cite{p4runtime-specification} extends P4 switches by an API to an SDN controller similar to OF.

In this paper, we consider MACsec to dynamically protect links between switches in SDN.
We propose to utilize an SDN controller for automated deployment of MACsec on SDN switches instead of relying on distributed protocols.
The SDN controller continuously monitors the topology of the network and sets up MACsec for detected links.
SDN switches implement the header structures and functionalities of MACsec, including encryption and decryption using AES-GCM.
OF only supports control plane programmability, i.e., MACsec data plane functionality cannot be implemented on the SDN switches.
Therefore, we propose a concept for MACsec in P4-based SDN and call it P4-MACsec.
P4 switches implement packet switching based on MAC addresses and MACsec, i.e., MACsec encryption, decryption, and integrity checks of packets.
For efficiency reasons, P4 switches are steered by a two-tier control plane.
Each P4 switch runs a local controller that connects to a central controller.
Functions of the control plane may be solely part of the local controller or part of both tiers.
The control plane implements MAC address learning for packet switching, a novel process for secure link discovery with encrypted LLDP packets, and automated deployment of MACsec.
To demonstrate the feasibility of P4-MACsec, we provide a prototype based on the BMv2 P4 software switch \cite{bmv2-github}.
We perform a functional validation of P4-MACsec in a Mininet testbed through experiments and investigate on TCP throughput and round-trip time (RTT) by conducting a performance evaluation.
We publish the source code of the prototype and all experiments on Github \cite{github}.
In addition, we report on experiences in implementing P4-MACsec for the NetFPGA SUME \cite{netfpga-website} platform.

The rest of the paper is structured as follows.
In \sect{macsec}, we review technical background and related work for IEEE 802.1AE (MACsec).
\sect{link-discovery} discusses technical background and related work on link discovery in SDN.
In \sect{p4}, we give an overview on P4.
\sect{concept} describes the architecture of P4-MACsec.
In \sect{implementation-mininet}, we describe the prototypical implementation of P4-MACsec with Mininet that is validated in \sect{validation}.
In \sect{evaluation}, we present a performance evaluation of the Mininet prototype.
In \sect{implementation-hardware}, we report on experiences in implementing P4-MACsec on the NetFPGA SUME platform.
\sect{conclusion} concludes this work.
  \section{MACsec: Foundations and Related Work}
\label{sec:macsec}

We give an overview of IEEE 802.1AE (MACsec) and explain how it protects the Ethernet layer.
We describe mechanisms for configuration and key management and review related work on the application of MACsec in SDN.

\subsection{Overview of MACsec}
IEEE 802.1AE \cite{8021AE_2006} introduces the media access control security (MACsec) protocol.
It provides point-to-point security between MACsec peers that are connected to the same local area network (LAN).
Examples are links between two switches or routers, links between switches or routers and hosts, and links between hosts.
MACsec ensures the integrity, confidentiality, and authenticity of Ethernet (IEEE 802) frames through applying symmetric encryption and cryptographic hash functions.
In addition, it provides replay protection and a key exchange protocol to ensure perfect forward secrecy, i.e., session keys are not affected by a compromised private key.

\begin{figure}[htp]
    \centering
    \includegraphics[width=\linewidth]{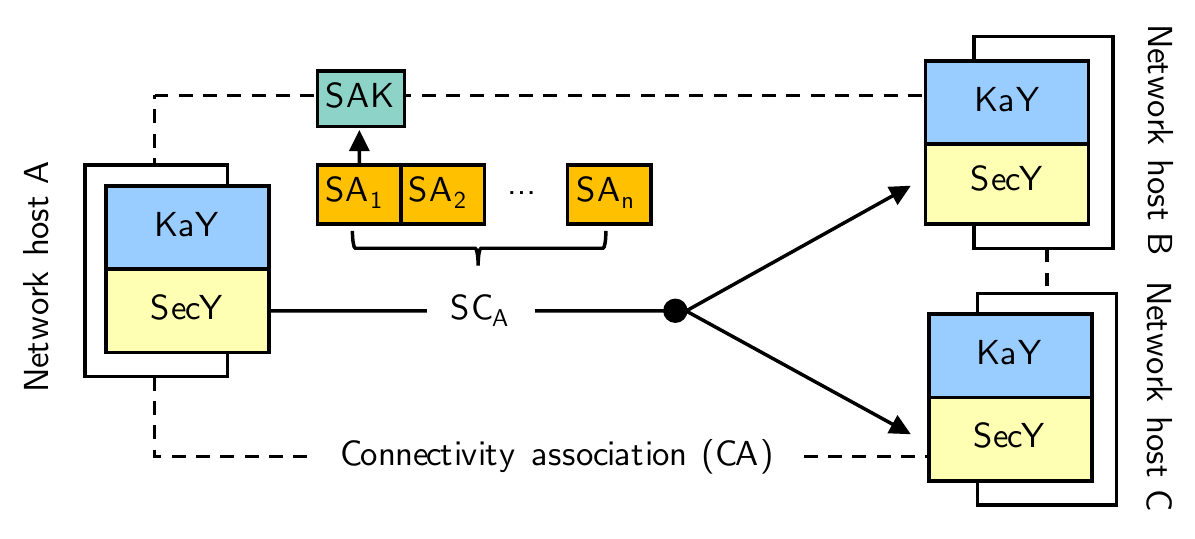}
    \caption{Secure communication between three stations in MACsec that are part of a connectivity association (CA). The unidirectional secure channel (SC) between the SecYs holds multiple security associations (SAs) each with an security association key (SAK) for encryption and decryption.}
    \label{fig:macsec-overview}
\end{figure}

\fig{macsec-overview} visualizes the principle and the core components of MACsec.
The network hosts A, B, and C are part of a LAN.
Each network host has a MAC security entity (SecY) and a MAC security key agreement entity (KaY).
The SecY provides secure MAC services over an insecure MAC service, i.e., it performs packet encryption and decryption.
The KaY discovers other KaYs in the LAN that participate in the same connectivity association (CA).
It ensures that all network hosts are mutually authenticated and authorized.
Afterwards, it creates and maintains secure channels (SCs) between the MACsec peers that are used by the SecY to transmit and receive network packets.
SCs are sender-specific, unidirectional, point-to-multipoint channels.
Each SC holds multiple secure associations (SAs) that have a secure association key (SAK) used for encrypting, decrypting, and authenticating packets.

MACsec leverages cipher suites for packet encryption, decryption, and authentication.
The standard defines the Advanced Encryption Standard in Galois/Counter mode (AES-GCM) with a block length of 128 bit (AES-GCM-128) as required cipher suite.
If only packet authentication but no encryption is configured, MACsec applies the Galois Message Authentication Code (GMAC).
Further specifications \cite{802.1AEbn-2011, 802.1AEbw-2013} add GCM-AES-256, GCM-AES-XPN-128, and GCM-AES-XPN-256, but other cipher suites that meet several requirements defined in the standard may also be applied. 

\begin{figure}[htp]
    \centering
    \includegraphics[width=.85\linewidth]{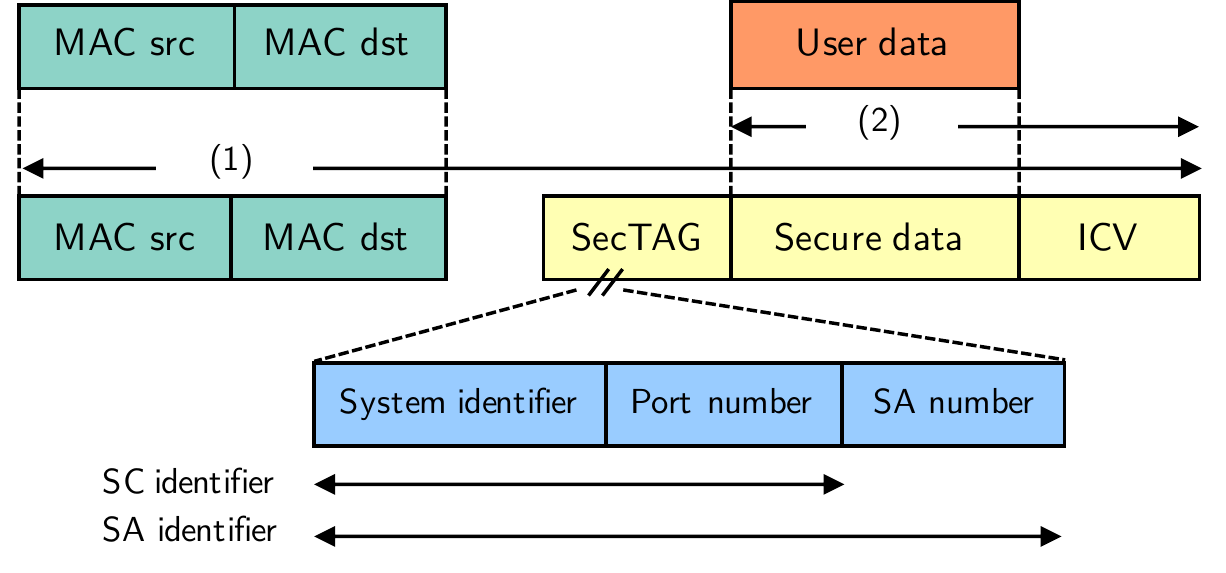}
    \caption{Packet structure of MACsec applied to an Ethernet packet. The MAC source and destination addresses of the MACsec packet are adopted from the original packet. The user data is transformed into a secure data block that is followed by the ICV calculated over the whole packet (1). The SecTAG includes among other things parameters to identify the SC and SA.}
    \label{fig:macsec-packet}
\end{figure}

\fig{macsec-packet} depicts the packet structure of MACsec.
The Ethernet source and destination addresses of the MACsec packet are adopted from the original Ethernet packet.
The secure data field either contains the encrypted user data of the original Ethernet packet or the user data in plaintext if only MACsec packet authentication is configured.
The integrity check value (ICV) field holds the result of a cryptographic hash function that is applied on the whole Ethernet packet (1) including all header fields.
The secure data and ICV is calculated by the chosen cypher suite (2).
The security tag (SecTAG) contains MACsec information, e.g., the SC or SA identifier to choose the corresponding SAK for packet encryption, decryption, or authentication.
SecYs store multiple SAs with SAKs for each SC.
When SAKs should be changed in rekeying, one SecY uses a new SA and signals the usage within the SecTAG so that the receiving SecY can also change to the new SA.

MACsec is supported by most managed access, distribution, and core switches from major manufacturers.
In addition, it is part of the Linux kernel since version 4.6 \cite{macsec-kernel} so that even switch-to-host or host-to-host links can be protected.

\subsection{MACsec Key Exchange Protocol}
\label{sec:macsec-mka}
The 802.1AE standard does not define processes for key management or establishment of CAs and SAs between KaYs on MACsec peers.
Therefore, network administrators are required to configure the CA affiliation and SAs with SAKs on every MACsec peer.
IEEE 802.1X-2010 \cite{8021X_2010} introduces the MACsec Key Agreement Protocol (MKA) for automated peer discovery and exchange of SA data.
With MKA, the initial CA affiliation and SA with SAK is derived from a connectivity association key (CAK).
CAKs are either defined as pre-shared secret, derived from a master session key of an EAP process, or distributed by a MKA key server.
Switches are required to implement additional functionalities to either exchange keying information via EAP with an AAA server or with an MKA key server.
Therefore, additional configuration effort is still required for each switch.

\subsection{Comparison to VPNs}
VPN technologies such as IPsec, OpenVPN, or WireGuard operate on Layer 3 or above.
MACsec operates on Layer 2 and therefore provides link security for any higher-layer protocol.
It applies point-to-point protection while VPNs aim at end-to-end protection.
On every switch, router, or host in the network, MACsec packets are decrypted at the ingress port so that control plane functions targeting Layer 2 to 7 can be still applied.
Access control lists (ACL) that provide filtering based on IP addresses are an example. 
Then, packets are encrypted again at the egress port.
MACsec is configured per Ethernet link so that administrators do not need to define additional policies for specific traffic to be encrypted.
On routers and switches, MACsec is implemented on the packet forwarding chips, i.e., packet encryption and decryption is performed in line rate.
In contrast, VPN technologies mostly encrypt and decrypt packets on ASICs that have limited bandwidth capacity.
According to \cite{cisco-macsec-innovation}, IPsec traffic typically cannot exceed 40 Gb/s of bidirectional traffic while MACsec encryption and decryption scales with line rate.

\subsection{Application of MACsec in SDN}
The authors of \cite{ChMi18} adopt MACsec to secure communication in vehicular networks between Linux-based electronic control units (ECUs).
An SDN controller is responsible for automated setup of MACsec between ECUs to provide an end-to-end protection for network traffic.
However, MACsec deployment is limited to the ECUs, i.e., MACsec deployment on SDN switches that connect the ECUs is not considered.
The authors of \cite{SzSa18} develop an intent-based multilayer orchestrator as application that interfaces an SDN controller.
It automatically deploys protection technologies such as IPsec or MACsec on legacy switches through different southbound protocols, e.g., OpenFlow, NETCONF, or RESTCONF.
However, MACsec deployment on SDN switches is not considered.
The authors of \cite{befl18} propose to implement MACsec for SDN but do not formulate any concrete approach.
The authors of \cite{VaKa16} discuss implementation experiences and design challenges for WAN overlays using SDN and propose MACsec as viable option to implement link layer encryption.
However, the presented implementation is limited to OpenVPN.
Automated configuration of MACsec on SDN switches is proposed, but not part of the presented implementation.
The authors of \cite{mohamed2015network} describe a mechanism for MACsec key distribution of particular MACsec flows to switches.
MACsec flows are end-to-end SCs that break up the point-to-point concept of the original standard.
They are realized by configuring MACsec keys only on both end peers, but not on the peers in between.
As prerequisite, all MACsec peers are expected to forward MACsec packets if no key for the received packet is found.

  \section{Link Discovery in SDN: Foundations and Related Work}
\label{sec:link-discovery}

Automated deployment of MACsec by a SDN controller requires a topology view that is maintained through topology monitoring with link discovery between SDN switches.
We give an overview on link discovery in SDN and describe the OpenFlow Discovery Protocol (OFDP).
We review related work on variants of the OFDP that are optimized regarding security, efficiency, and applicability in hybrid SDN networks.

\subsection{Topology Monitoring and Link Discovery in SDN}
Topology monitoring in SDN maintains a network map on the SDN controller that consists of SDN switches and links in between.
In contrast to legacy networks, topology monitoring in SDN can be limited to link discovery.
In OpenFlow (OF), SDN switches establish a connection to a pre-configured SDN controller during the start.
The SDN controller receives information about the SDN switch, e.g., a list of all physical ports, within the OF handshake at connection setup.
With the P4 Runtime API, the SDN controller may connect to P4 switches during the start.
In both cases, the SDN controller already identified all SDN switches so that only links need to be detected.

\subsection{OpenFlow Discovery Protocol (OFDP)}
\label{sec:ofdp}
\begin{figure}[htp]
    \centering
    \includegraphics[width=\linewidth]{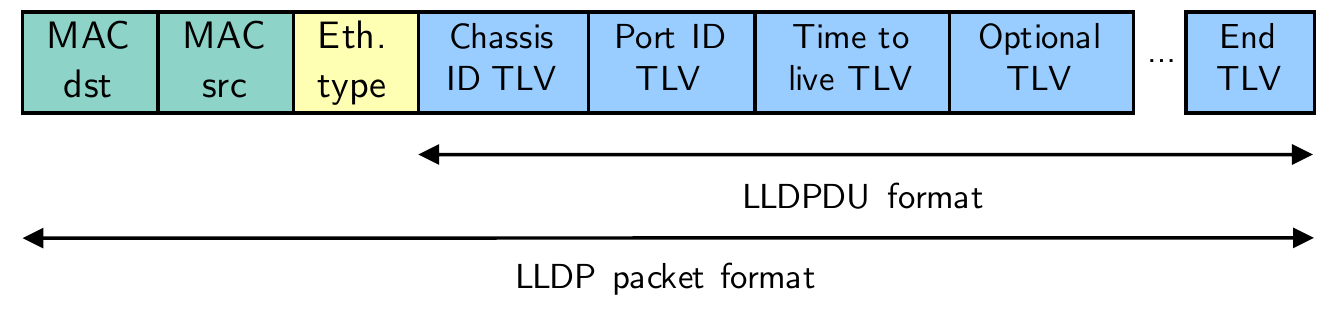}
    \caption{Format of LLDP packet and LLDPDU.}
    \label{fig:lldp-packet-format}
\end{figure}

The OpenFlow Discovery Protocol (OFDP) was the first de-facto standard for link discovery in SDN.
It leverages the Link Layer Discovery Protocol (LLDP) \cite{8021AB_2009}, the most widely used protocol for link detection in legacy networks.
The Cisco Discovery Protocol (CDP) \cite{cdp} is a proprietary alternative but less widely used.
LLDP advertisements include information about the identity of a host, its capabilities, and its current status.
LLDP protocol data units (PDUs) are periodically sent as payload of Ethernet frames with a multicast receiver address and the EtherType 0x88cc.
\fig{lldp-packet-format} depicts their structure.
The PDUs may contain various type-length value (TLV) blocks, the standard defines three required TLV blocks.
First, the Chassis ID TLV identifies the sending host, e.g., by its MAC address.
Second, the Port ID TLV identifies the sender's port, e.g., its physical port number.
Last, the Time-to-Live TLV defines the time validity of the information. 
Optional TLV blocks, e.g., the system's name defined by the administrator, and custom TLVs may be used as well.
Network hosts that implement LLDP can receive, but not request LLDP information.
Legacy switches periodically send out LLDP packets on each active port as described before.
The packets are received, processed, and dropped by neighbouring LLDP agents on switches.
They store the received information in the management information bases (MIBs) that can be queried by SNMP.

\begin{figure}[htp]
    \centering
    \includegraphics[width=0.9\linewidth]{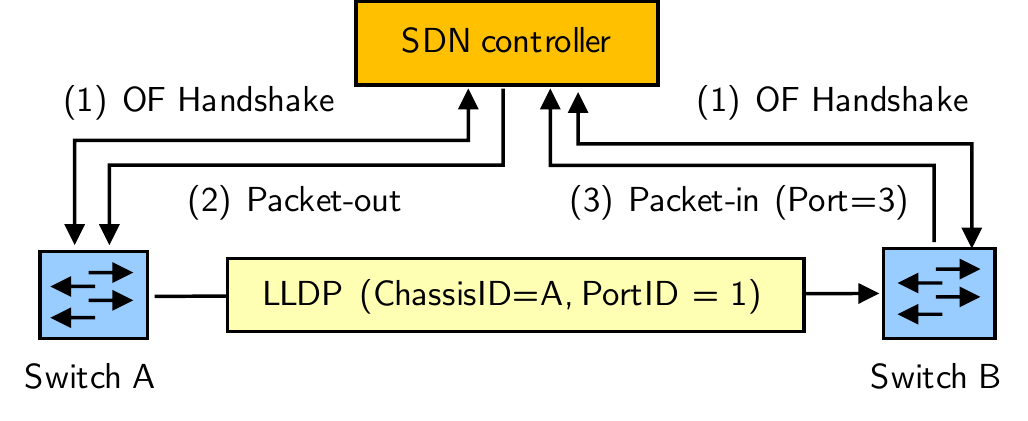}
    \caption{Link discovery in SDN with OFDP. The SDN controller learns about the SDN switch within the OF handshake (1). Afterwards, it sends out an LLDP packet on a particular port of an SDN switch via a packet-out message (2). Another SDN switch that receives the LLDP packet forwards it back to the SDN controller using a packet-in message (3).}
    \label{fig:ofdp-discovery}
\end{figure}

The OFDP leverages LLDP as introduced before but delegates all functionalities to the SDN controller.
It uses packet-out messages to send out created network packets over particular ports of a SDN switch and packet-in messages to receive packets from the SDN switch that match specific criterias, e.g., an LLDP EtherType.
\fig{ofdp-discovery} depicts the process of link discovery with OFDP.
First, the SDN controller learns about the switch identity and its ports within the OF handshake (1).
Afterwards, it creates dedicated LLDP packets for all ports of a switch that are sent out via packet-out messages (2).
For incoming LLDP packets, the OF switches are configured to forward any LLDP packet as packet-in message to the SDN controller (3).
The packet-in message includes the LLDP packet with the Chassis and Port ID of the sender along the identity and the ingress port of the receiving SDN switch.
By repeating this process for each port on each switch, the SDN controller performs link discovery.

\subsection{Optimized Variants of OFDP}
\label{sec:optimized-variants-ofdp}

We review related work on optimized variants of OFDP that can be subdivided into publications investigating the security of OFDP, efficiency of OFDP, and applicability of OFDP in hybrid networks.

\subsubsection{Security of OFDP}
The authors of \cite{AlPo15}, \cite{AlPo18}, \cite{AzTh17}, and \cite{NgYo17} show that OFDP is vulnerable to spoofing attacks.
Injected LLDP control messages may create fake links that redirect traffic to the host of an attacker.
The authors of \cite{AzTh17} show that OFDP is additionally vulnerable to controller fingerprinting, switch fingerprinting, and LLDP flooding attacks.
The authors of \cite{NgYo17} show that OFDP is vulnerable to replay attacks of LLDP packets that result in incorrect link information of the topology.
As improvement, the authors of \cite{AlPo15} and \cite{AlPo18} propose to add a message authentication code (MAC) and a message identifier to each packet to provide authentication, packet integrity, and to prevent replay attacks.
sOFTDP \cite{AzBo18} encrypts LLDP packets to further prevent fingerprinting attacks.

\subsubsection{Efficiency of OFDP}
The authors of \cite{AzTh17} and \cite{RoAl18} show that OFDP results in too many packet-out messages as the SDN controller has to create and send out one message for every port on each SDN switch.
As an improvement, the authors of \cite{onos-link-discovery} propose to apply LLDP with Port IDs set to zero.
The SDN controller creates one LLDP packet for every SDN switch that are configured to output the LLDP packet on all ports.
This process is repeated for all SDN switches.
Adjacent SDN switches are configured to forward received LLDP packets to the SDN controller so that it learns about the unidirectional link.
The authors of \cite{PaPo14} propose to reduce the number of packet-out messages through rewriting LLDP packets on the SDN switch.
sOFTDP \cite{AzBo18} introduces several mechanisms to shift large parts of link discovery back to the SDN switch.
It adds liveliness detection mechanisms for switch ports and memorizes topology information locally on the SDN switch that asynchronously notifies the SDN controller about specific events.
The authors of \cite{RoAl18} propose the Tree Exploration Discovery Protocol.
SDN controllers create and send out specific frames flooded in the network that explore its topology.
However, all concepts that shift functionality back to the SDN switches require extensive functional changes on the fixed-function data plane of typical SDN switches.

\subsubsection{Applicability of OFDP in Hybrid Networks}
Hybrid networks consist of SDN and non-SDN switches.
OFDP can be only applied to detect links in networks that consist of SDN switches.
Legacy switches that may connect SDN switches process and discard received LLDP packets.
The Broadcast Domain Discovery Protocol (BDDP) is a non-standardized approach that is implemented by several SDN controllers \cite{opendaylight,floodlight,onos}.
BDDP messages adapt the LLDP packet structure but use a broadcast Ethernet address instead of a multicast Ethernet address and the custom EtherType 0x8999.
SDN switches are programmed to forward received BDDP packets to the SDN controller, just as with LLDP.
Legacy switches flood the packet through all ports because of the broadcast address.
They relay BDDP packets so that links will appear as single hops no matter how many legacy devices are on the path between two SDN switches.
However, the authors of \cite{OcCe15} show that the usage of broadcast packets leads to inefficient and excessive utilization of network resources.
The authors propose a two-phase process for topology detection.
First, the SDN controller performs link discovery using OFDP as described before.
Afterwards, it outputs BDDP packets on any active port for which it does not detect a direct link to another SDN switch via LLDP.
This way, the SDN controller detects direct links via LLDP and indirect links via BDDP.
  \section{P4: Foundations}
\label{sec:p4}

We briefly overview P4 with its core components for programming a P4 switch, describe the P4 Runtime, and present examples for P4 software and hardware targets.

\subsection{Overview}
P4 is a domain-specific language for programmable data planes of network switches.
It offers high-level constructs that are optimized for specifying the forwarding behaviour of a switch.
P4 was first published in 2014 \cite{p4-paper}.
Today, the specification and development takes place in the non-profit P4 Language Consortium \cite{p4-website} with over 110 members from industry and academia.
$P4_{16}$ \cite{p4-specification} is the latest version of the language specification, the source code of all related components is available under a Apache license.

\begin{figure}[htp]
    \centering
    \includegraphics[width=0.98\linewidth]{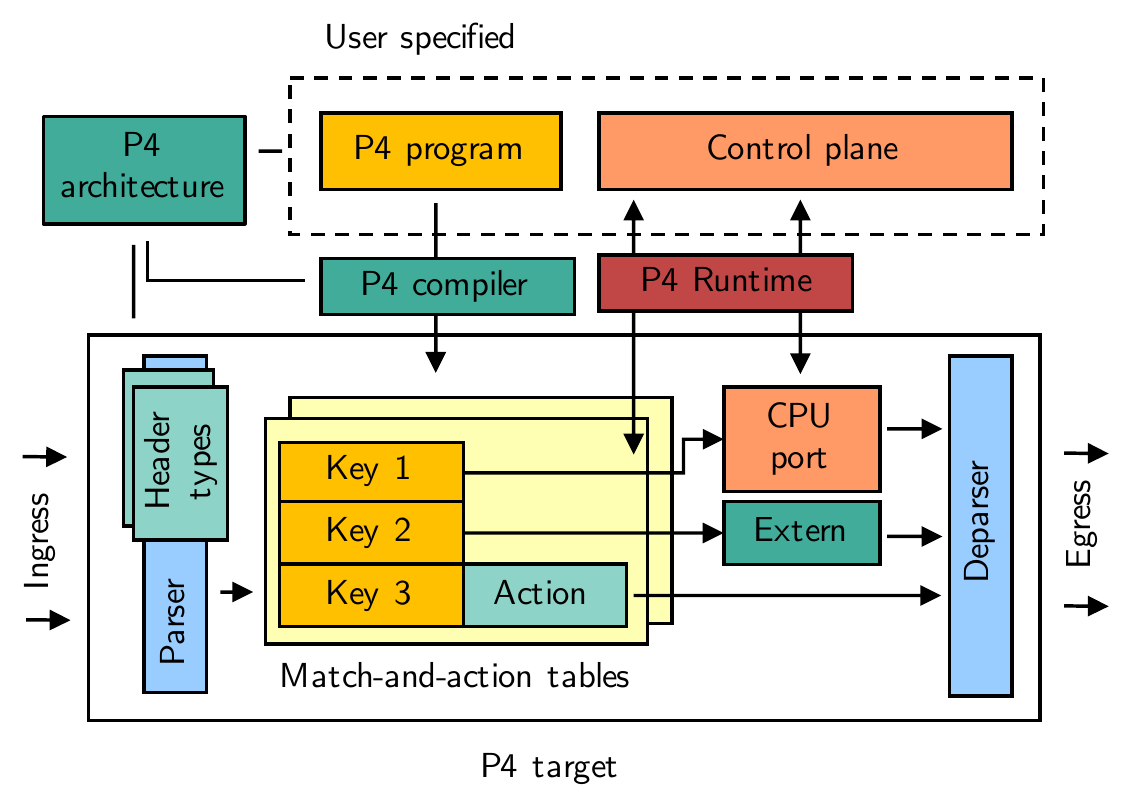}
    \caption{Concept and core components of $P4_{16}$.}
    \label{fig:p4}
\end{figure}

\fig{p4} depicts P4's core concept and components.
P4 programs contain the whole forwarding behaviour of switch.
They are formulated for a particular P4 architecture describing the programming model of a switch.
P4 targets are software or hardware switches that implement a specific P4 architecture.
Target-specific P4 compilers generate binary code from P4 programs that can be loaded on the P4 target.

\subsection{Core Components for Programming a P4 Switch}
In the following, we briefly describe the core components of the P4 programming language following the specification of $P4_{16}$.
All components are shown in \fig{p4}.
First, P4 header types describe the format of packet headers through an ordered collection of base types.
For example, an Ethernet header is described by bit vectors for the MAC source address, the MAC destination address, and the EtherType.
Second, P4 parsers are state machines that extract packet data through applying predefined sequences where data is identified and extracted based on P4 header types.
For instance, the value of a parsed EtherType field of a packet determines the following extraction state which could be LLDP, MACsec, or IP. 
Third, P4 tables are match-and-action structures mapping user-defined keys to particular P4 actions that may manipulate packet data.
Fourth, P4 externs are functions provided by a P4 target that can be used within P4 programs.
P4 externs have a defined interface with a set of methods that can be used in the P4 programm.
An example would be a function that calculates checksums for given chunks of data.
Last, the P4 deparser assembles the headers back into a well-formed network packet that can be sent out via an egress port of the switch.

\subsection{P4 Runtime}
The P4 Runtime framework provides an API for controlling P4 targets.
Its operation is visualized in \fig{p4}.
The P4 Runtime features the manipulation of match-and-action tables through the control plane.
In addition, it provides a CPU port for sending out and receiving packets similar to the packet-in and packet-out mechanism known from OpenFlow.
P4 Runtime leverages gRPC \cite{grpc} that is based on HTTP/2 and protocol buffer \cite{protocol-buffers} data structures.
The connection between P4 switches and the control plane can be secured through TLS with optional client and server certificates for mutual authentication.

\subsection{P4 Software \& Hardware Targets}
The BMv2 \cite{bmv2-github} is the most widely-used P4 software switch that features multiple P4 targets.
Examples are a P4 switch with the P4 Runtime API (simple\_switch\_grpc) or a P4 switch implementing the Protocol Independent Switch Architecture (PISA).
The NetFPGA-SUME \cite{netfpga-ieee} or the Netcope NFB-200G2QL \cite{netcope} are P4 targets that are based on Field Programmable Gate Array (FPGA) platforms.
Ethernet switches featuring the Barefoot Tofino ASIC \cite{tofino-asic} are the most widely-used P4 switches nowadays.
The Barefoot Tofino ASIC implements the PISA architecture and offers 12.8 Tb/s of packet throughput in its latest version.
It is part of Ethernet and Whitebox switches that also feature a general-purpose computing unit with a x86 CPU, RAM, and a SSD that runs a Linux-based operating system.
An example is the Edgecore Wedge 100BF-32X \cite{wedge-datasheet} Whitebox switch.
  \section{P4-MACsec}
\label{sec:concept}

In this section, we describe P4-MACsec.
We review its architecture and outline the three functional parts, packet switching with MAC address learning, secure link discovery, and automated deployment of MACsec, in detail.

\subsection{Overview}
\begin{figure}[htp]
  \centering
  \includegraphics[width=.98\linewidth]{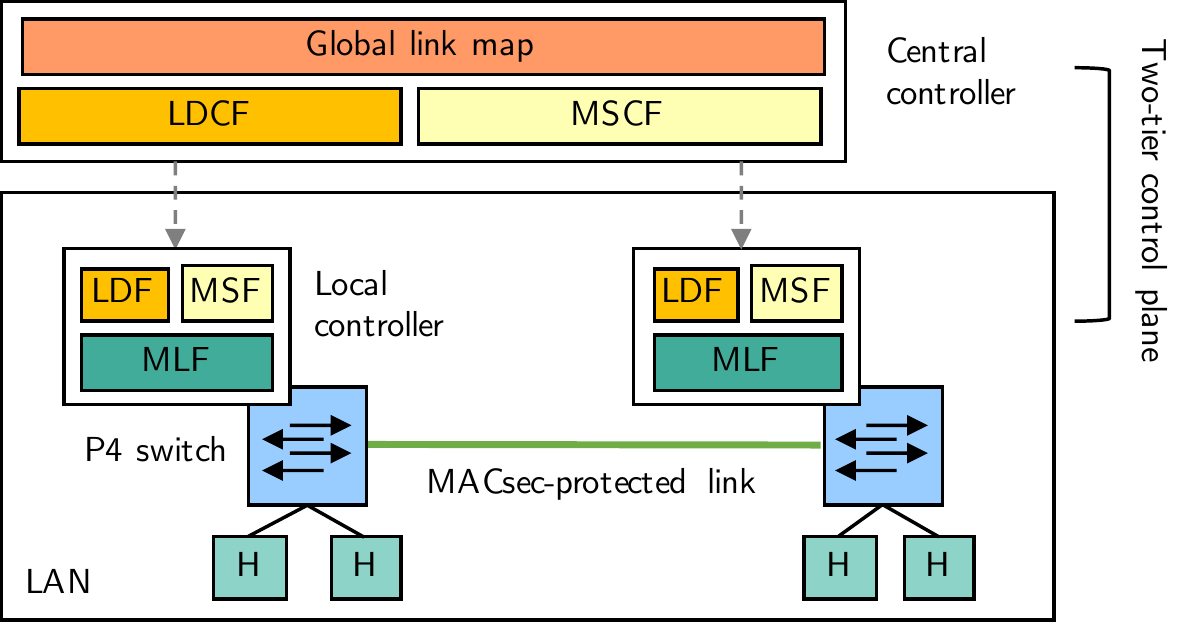}
  \caption{P4-MACsec in LAN with P4 switches. The local controllers consist of the MAC address learning function (MLF), link discovery function (LDF), and MACsec function (MSF). They communicate with the central controller that runs the link discovery controller function (LDCF), MACsec controller function (MSCF), and the global link map.}
  \label{fig:overview}
\end{figure}

\fig{overview} depicts the concept of P4-MACsec.
It consists of a LAN with P4 switches that are steered by an SDN control plane.
P4 switches connect network hosts and implement the data plane functionality of the three functional parts.
P4-MACsec features a \emph{two-tier control plane structure}.
Each P4 switch runs a local controller that connects to a central controller.
The two-tier control plane structure allows functions to be shifted to the local controller or split up into two parts, one running on the local controller and one part running on the central controller.
This reduces traffic in the management network, load on the central controller, and latency from forwarding packets between the SDN switches and the SDN controller.
The concept of hierarchical SDN control is not novel but part of several SDN control plane architectures (e.g., \cite{orion, kandoo, zerosdn}).
All P4 targets are coupled with general-purpose computing capacities that may run a local controller on the same device.
In contrast to OpenFlow or other SDN architectures, the two-tier control plane structure therefore does not introduce additional computing nodes that may imply a further risk of failout.
The authors of \cite{BaSo18} provide an overview on hierarchical and distributed SDN control planes and discuss its specifics.
Although we see many advantages in the two-tier control plane architecture, P4-MACsec can be implemented with a one-tier control plane structure featuring a central SDN controller as well.

We design the three functional parts as follows.
First, \emph{MAC address learning for packet switching} ensures that P4 switches forward Ethernet packets to provide Layer 2 connectivity for all network hosts.
We implement it as MAC address learning function (MLF) that is exclusively part of the local controller and describe its details in \sect{ethernet-forwarding}.
Second, \emph{secure link discovery} detects and monitors the link topology of the network using 
LLDP protocol data units (PDUs) that are protected with AES-GCM.
We implement it as two-tier control plane function.
It consists of link discovery functions (LDFs) running on the local controllers that inform the link discovery controller function (LDCF) running on the central controller about their local link view.
The LDCF composes the global link map that is the basis for automated deployment of MACsec.
We describe the details of this functional part in \sect{secure-link-discovery}.
Third, \emph{automated deployment of MACsec} dynamically creates, sets up, and maintains SCs on all switch-to-switch links from the global link map.
We implement it as two-tier control plane function with decentral MACsec functions (MSF) that receive configuration from a MACsec controller function (MSCF) running on the central controller.
We describe its details in \sect{automated-deployment-macsec}.

\subsection{Ethernet Packet Switching with MAC Address Learning}
\label{sec:ethernet-forwarding}
Although MAC address learning is a typical example for a local switch function, it cannot be solely implemented on the data plane of a P4 switch.
Our proposed architecture for that functional part consists of two components.
First, a MAC address learning function (MLF) that runs on the local controller.
Second, the data plane implementation for MAC address learning with the MLF and packet switching.

\begin{figure}[htp]
  \centering
  \includegraphics[width=0.98\linewidth]{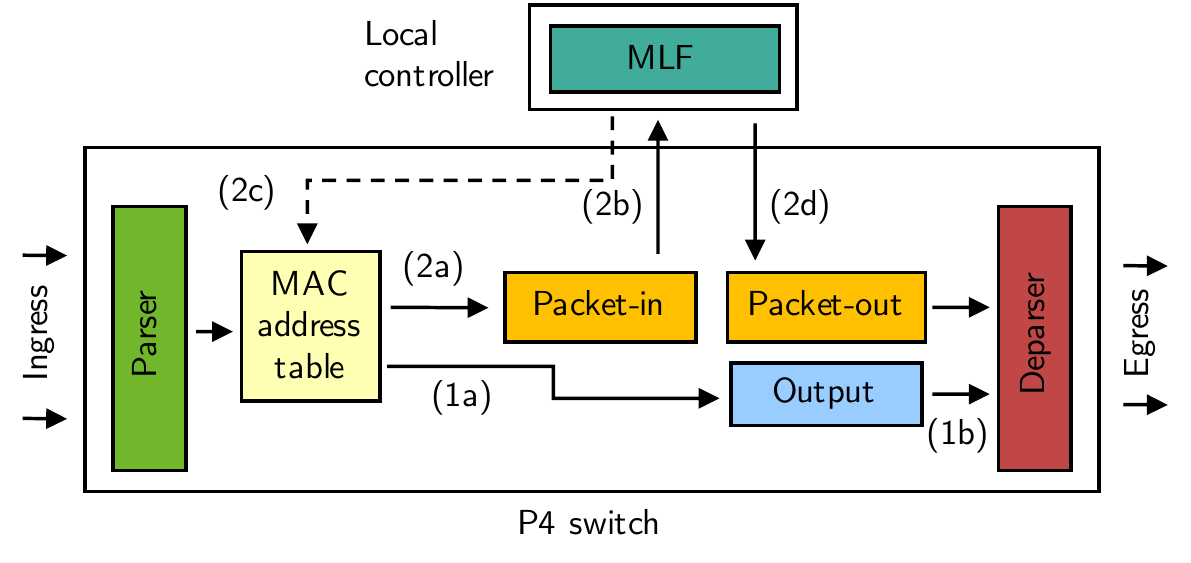}
  \caption{Process of forwarding Ethernet packets on the P4 processing pipeline. In case of missing entries in the MAC table, the processing pipeline issues MAC address learning that is performed by the MAC address learning function (MLF) running on the local controller.}
  \label{fig:mac-learning}
\end{figure}

\fig{mac-learning} visualizes the process with all interactions between both components.
When the P4 switch receives an Ethernet packet, it first checks if the source and destination MAC address of the Ethernet packet are already part of the MAC address table.
If the MAC address table has entries for both (1a), the switch forwards the packet on the port specified for the destination MAC address (1b).
If the MAC table yields no match for both addresses (2a), the switch forwards the Ethernet packet to the MLF running on the local controller as packet-in message (2b).
The MLF first checks if the MAC source address and the ingress port is already part of the MAC address table.
If not, the MLF updates the MAC address table (2c).
Afterwards, the MLF floods the Ethernet packet on all ports except the ingress port where it received the packet through the packet-out function (2d).

\subsection{Secure Link Discovery}
\label{sec:secure-link-discovery}
The functional part of secure link discovery consists of three components.
First, link discovery functions (LDFs) running on local controllers that perform link detection and monitoring.
Second, the link discovery controller function (LDCF) running on the central controller that composes the global link map from local link information received from the LDFs.
Third, the data plane implementation for receiving and sending out LLDP packets via packet-in and packet-out messages.

\begin{figure}[htp]
  \centering
  \includegraphics[width=0.95\linewidth]{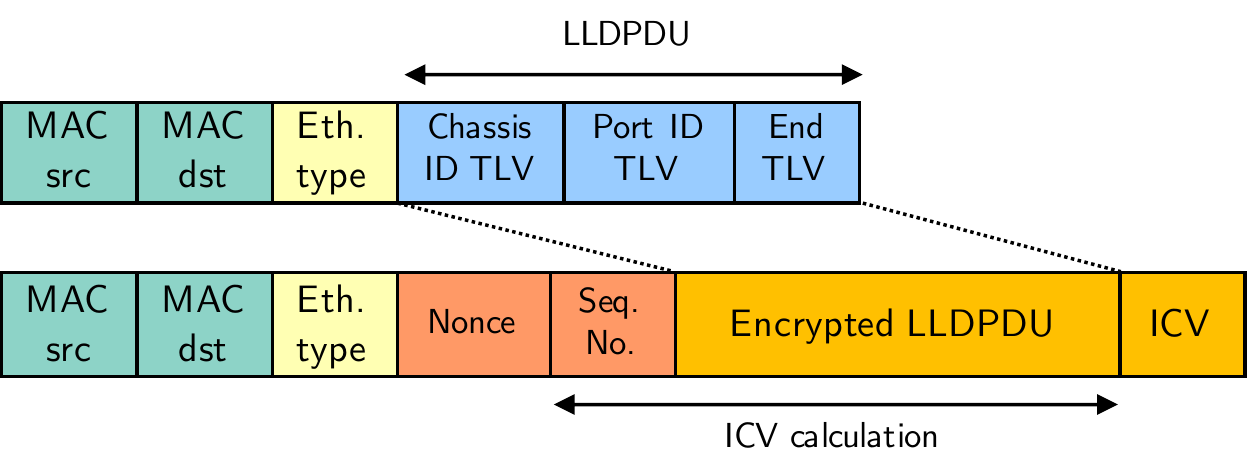}
  \caption{Proposed format for encrypting LLDPDUs with AES-GCM. The EtherType from the original packet is followed by a Nonce that is used in AES-GCM encryption and decryption. A sequence number protects against replay attacks, a ICV holds a cryptographic checksum for authentication.}
  \label{fig:encrypted-lldp}
\end{figure}

As novelty, we propose to create, encrypt, and decrypt LLDPDUs on the LDF using AES-GCM with a common encryption key.
We additionally introduce sequence numbers for LLDPDUs to defend them against replay attacks.
\fig{encrypted-lldp} visualizes our proposed format of LLDPDUs in comparison to the original format of LLDP packets.
As in legacy LLDP, the MAC source address is set to the MAC address of the P4 switch.
The MAC destination address and the EtherType are set to the LLDP defaults as introduced in \sect{ofdp}.
LLDPDUs consists of three TLVs.
The Chassis ID TLV contains the identity of the switch as defined by the network administrator, the Port ID TLV contains the number of the physical port, the End TLV marks the end of the LLDPDU.
The common encryption key is installed and frequently updated by the LDCF on all LDFs.
In addition, AES-GCM uses a 12 byte random number as nonce that is re-generated for each LLDPDU leaving a particular port.
It is part of the packet header following the EtherType.
The 4 Byte sequence number as protection against replay attacks is initialized with the LDF bootup timestamp and incremented with each packet sent out.
The receiving LDF holds a sequence number counter for every physical port that is incremented with any received packet.
The sequence number is part of the packet authentication of AES-GCM that is applied to the sequence number and on the LLDPDU.
Its result, the authentication tag, is stored within the ICV field following the encrypted LLDPDU.
Our approach is similar to \cite{AzBo18} and protects against all attacks that were discussed in \sect{optimized-variants-ofdp}.

\begin{figure}[htp]
  \centering
  \includegraphics[width=\linewidth]{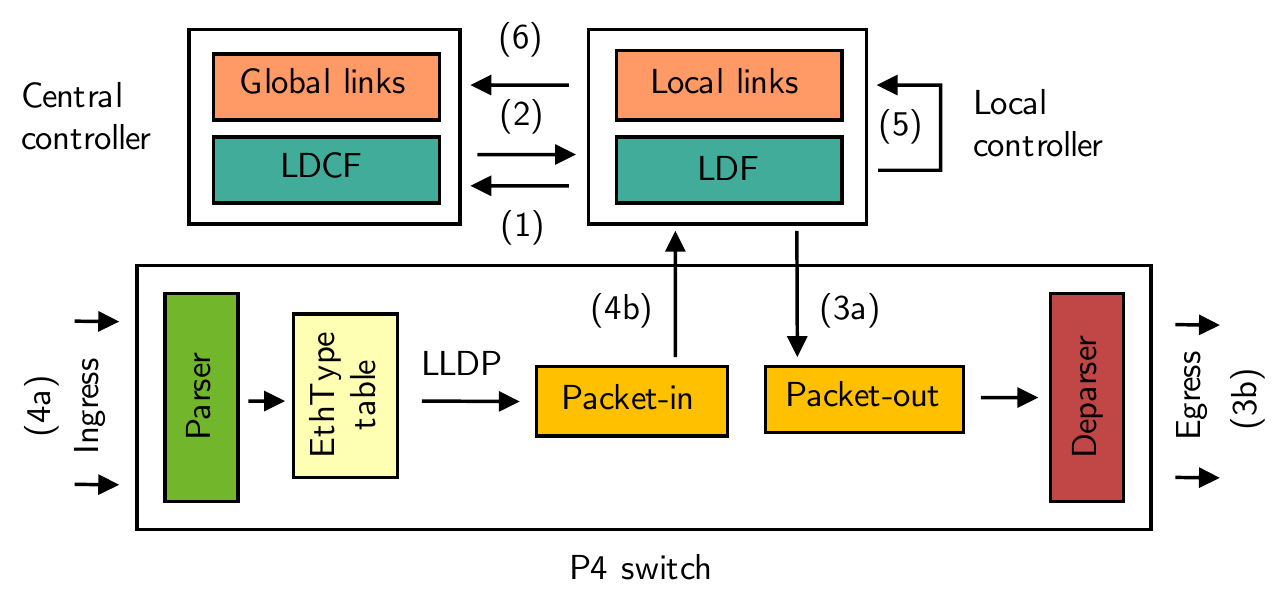}
  \caption{Process of secure link discovery using the local link discovery functions (LDF), the link discovery controller function (LDCF), and the processing pipeline on the P4 switch.}
  \label{fig:ldf-ldcf}
\end{figure}

\fig{ldf-ldcf} visualizes the process of secure link discovery with all interactions between the three components.
At startup, the LDF initiates a connection to the LDCF through a preconfigured IP address or FQDN (1).
The LDCF installs the common key that is used for encrypting and decrypting LLDP packets with AES-GCM and instructs the LDF to start secure link discovery (2).
The LDF generates and transmits encrypted LLDP packets for all active port via packet-out messages to the P4 switch (3a) which then outputs the received packets on the specified ports (3b).
As all other P4 switches received the same instruction to start link discovery, the P4 switch now receives encrypted LLDP packets from other P4 switches (4a) that it sends as packet-in messages to the LDF (4b).
It performs decryption and extracts the Chassis and Port ID of the distant switch from the LLDPDU along the physical port number of the packet-in message to update the map of local links (5).
Finally, all changes of the local link map are sent to the global link map on the central controller (6).
The global link map consists of bidrectional link in the form of two tuples that indicate the identity of the P4 switch and the physical port of the link as $(\textrm{Switch}_{\textrm{A}}, \textrm{Port}_{\textrm{A}}) \rightarrow (\textrm{Switch}_{\textrm{B}}, \textrm{Port}_{\textrm{B}})$.
Link topology can alter at any time, e.g., when links between switches are added or when cables break.
Therefore, link discovery is executed whenever the LDF running on the local controller receives status messages from its assigned switches, e.g., due to a port-down notification when a cable breaks.
Link discovery is additionally performed after a fixed time interval of $\textrm{30\ seconds}$ so that security can be sustained even if status messages from the P4 switches get lost.

In comparison to other approaches presented in \sect{optimized-variants-ofdp}, our proposal does not require modifications of the SDN switches.
All required mechanisms are implemented as control plane functions that rely on packet-in and packet-out mechanisms as offered by the CPU port in P4 or the packet-in and packet-out messages of OpenFlow.

\subsection{Automated Deployment of MACsec}
\label{sec:automated-deployment-macsec}
Automated deployment of MACsec consists of three parts.
First, a MACsec controller function (MSCF) that creates two unidirectional SCs for each link of the global link map.
Second, a MACsec function (MSF) that runs on the local controller.
It receives configuration data from the MSCF and sets up the SCs on the P4 switch.
Third, the data plane implementation of MACsec that consists of the P4 processing pipeline and implementations of the MACsec validate and protect functions that can be used as P4 externs.

\begin{figure}[htp]
  \centering
  \includegraphics[width=\linewidth]{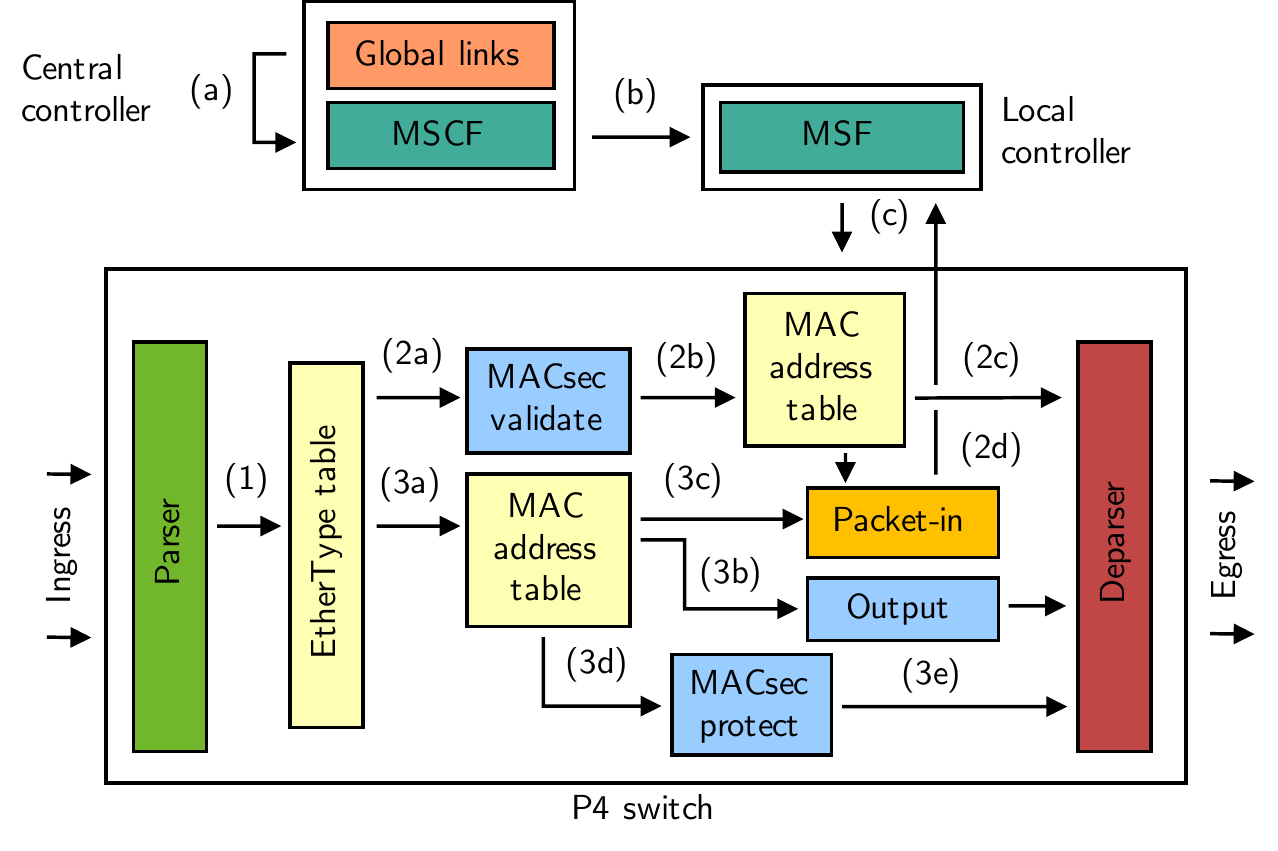}
  \caption{Process of automated MACsec deployment with the MACsec function (MSF), MACsec controller function (MSCF, and the processing pipeline of the P4 switch.)}
  \label{fig:processing-pipeline}
\end{figure}

\fig{processing-pipeline} visualizes the automated deployment of MACsec with all interactions between the three components.
We later describe the details of the MACsec validate and protect function.
The P4 processing pipeline is an extension of the Ethernet packet forwarding pipeline that was introduced in \sect{ethernet-forwarding}.
At start, the MSCF creates and maintains MACsec secure channels (SCs) based on the global link map (a).
It passes SC configuration data to the MSF (b) which updates various P4 tables in the processing pipeline (c).
For the P4 processing pipeline, it sets a MACsec flag to entries of the MAC address table if a SC for that particular link exists.
Then, the data plane processing for packets works as follows.
First, ingress packets are matched in an EtherType table (1).
Ethernet packets matching the MACsec EtherType are forwarded to the MACsec validate function (2a).
It validates its authenticity, optionally decrypts the secure data, and returns an Ethernet packet.
Afterwards, the processing pipeline continues with Ethernet packet forwarding, i.e., it consults the MAC address table (2b), outputs the packet in case of a match for both source and destination MAC address (2c), or sends the packet as packet-in message to the MLF on the local controller otherwise (2d).
Ethernet packets matching other EtherTypes are forwarded to the MAC address table (3a).
If the MAC address table holds no flag for an SC for the particular destination MAC address, the packets are either sent out (3b) or passed to the MLF (3c) as explained in \sect{ethernet-forwarding}.
If the MAC table yields a match for both MAC addresses of the packet and an SC flag, it forwards the packet to the MACsec protect function (3d).
The MACsec protect function responds with a MACsec packet that can be sent out (3e) via the egress of the processing pipeline.

\begin{figure}[htp]
  \centering
  \includegraphics[width=\linewidth]{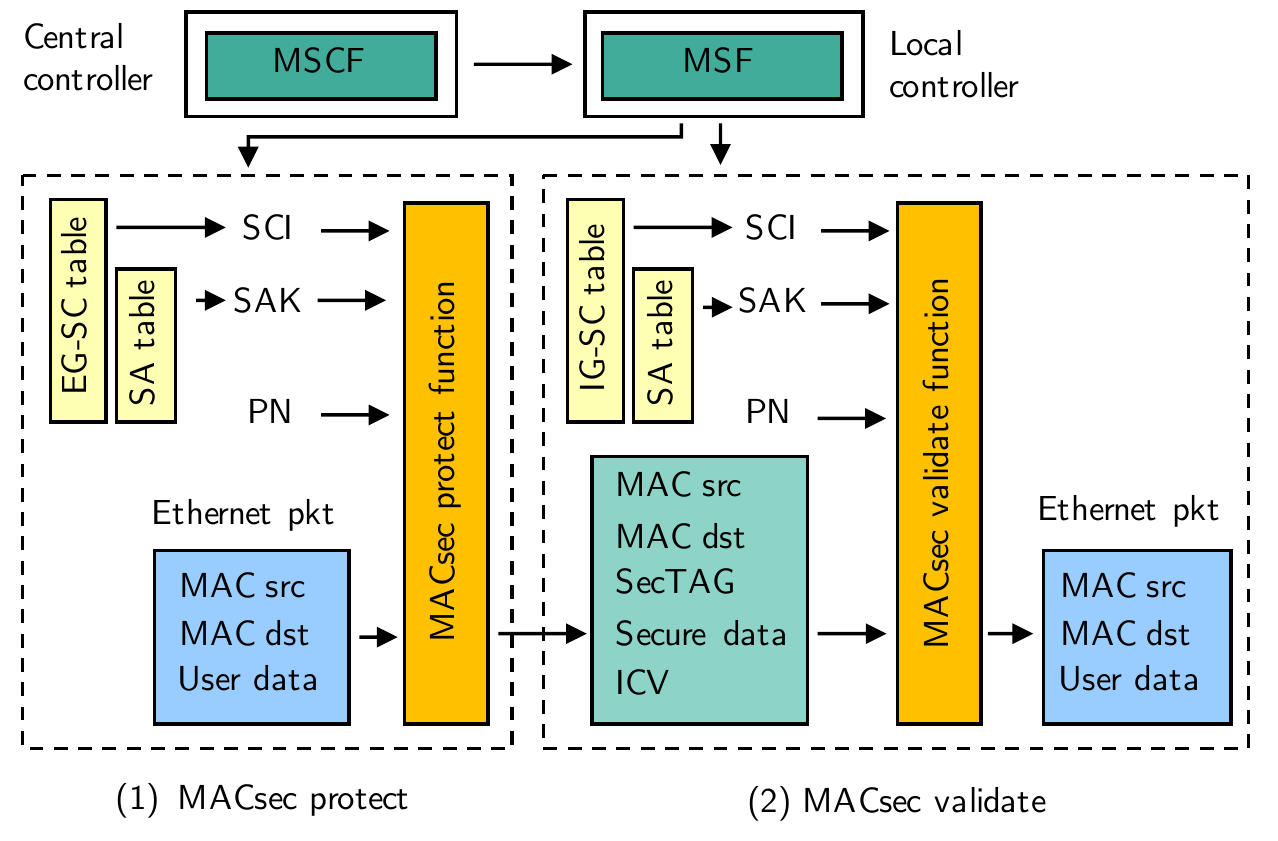}
  \caption{MACsec protect (1) and MACsec validate (2) functions implemented as P4 externs within the P4 processing pipeline.}
  \label{fig:macsec-protect-validate}
\end{figure}

\fig{macsec-protect-validate} visualizes the MACsec protect and validate function.
As SCs are unidirectional, the P4 switch has a ingress MACsec SC (IG-SC) table and a egress MACsec SC (EG-SC) table that maps secure channel identifiers (SCIs) to security associations (SAs).
SAKs are part of the SA table that holds SAs for both, ingress and egress SCs.
The MACsec protect function (1) either encrypts or authenticates Ethernet payloads.
It is applied to Ethernet packets if the MAC address table in the Ethernet packet forwarding pipeline has a flag set for the particular physical port.
Within the protect function, an egress secure channel (EG-SC) table maps the egress port number as SCI to security association identifiers (SAIs).
The SA table holds the security association keys (SAK) to be used for the protect function.
The MACsec protect function receives the SAK and SCI from the SA table, the packet number from a packet counter as part of the switch, and the Ethernet packet.
The AES-GCM cipher is initialized with a concatenation of the SCI and packet number as initialization vector.
Afterwards, the EtherType and the payload are concatenated and encrypted.
The MACsec protect function creates a new Ethernet packet with the MAC source and MAC destination address of the Ethernet packet, the MACsec EtherType, the SecTAG, the secure data, and the ICV.
The MACsec validate function (2) works in a similar manner.
Again, an ingress secure channel (IG-SC) table maps SCIs to SAIs.
The MACsec validate function receives the SAK and SCI from the IG-SC and SA table, the packet number from a packet counter, and the MACsec packet.
It then checks the integrity and optionally decrypts the packet.
It returns an Ethernet packet with the original Ethernet header and payload to the processing pipeline where forwarding and MAC address learning occurs as described before.

The MSCF creates and maintains MACsec SCs whenever the global link map changes.
SCs exist until the corresponding link is deleted, e.g., in case of a link failure, the corresponding SC with all its SAs is deleted.
If a new link is detected, the MSCF creates a new MACsec SC with SAs.
In addition, MACsec SCs and SAs are renewed on a regular basis.
Administrators define timeouts for encryption keys, the MSCF generates and installs new SAs on the P4 switches after the defined time interval through the MSF.
Whenever SCs are created, changed, or deleted, configuration data and SAs are passed to the MSF that programs the P4 switch through writing in the EG-SC, IG-SC, and SA table.
  \section{Prototypical Implementation with Mininet}
\label{sec:implementation-mininet}

In the following, we describe a prototypical implementation of P4-MACsec.
We review the Mininet testbed environment and describe the three components of P4-MACsec in detail.

\subsection{Testbed Environment}
We use the Mininet \cite{mininet} network emulator to build the testbed environment for the prototypical implementation.
We leverage the BMv2 P4 software switch \cite{bmv2} for implementing the P4 switch and run the local controllers and the central controller as Python applications.
For testing purposes, we additionally run Mininet network hosts that are connected to the P4 switches.
All testbed components are executed within a KVM/QEMU virtual machine (VM) that runs Ubuntu 16.04. with 4 CPU cores and 4GB RAM.
The hypervisor host features an Intel Core i5 8250U CPU, 16GB RAM, and an SSD.

\subsection{P4 Switch}
We extend the simple\_switch\_grpc \cite{bmv2-grpc} P4 target of the BMv2 P4 software switch \cite{bmv2} to later run our P4 program that describes the data plane functions.
\fig{implementation} depicts its parts.
First, we implement the MACsec protect and validate functions as P4 externs within the simple\_switch\_grpc P4 target.
We program the extensions in C++ and leverage the envelope (EVP) interface of OpenSSL \cite{openssl} to apply AES-GCM for encryption, decryption, and packet authentication.
Both functions can be used as P4 externs within the P4 processing pipeline.
When accessing the functions from the P4 processing pipeline, packet header data is exchanged using P4 attributes where packet payload data can be accessed directly.
Second, we implement an interface to the local controller.
It leverages the P4 runtime API via gRPC and allows the local controller to modify entries of the P4 tables in the processing pipeline.
In addition, it holds the CPU port interface that provides packet-in and packet-out messages as known from OpenFlow.
Packets sent from the P4 pipeline to the CPU port are forwarded to the local controller, packets received from the CPU port are injected into the P4 processing pipeline.
The data plane functions of P4-MACsec described as P4 processing pipeline in \sect{concept} are implemented as P4\_16 program using known P4 constructs as introduced in \sect{p4}.
The P4 program then is executed on the modified simple\_switch\_grpc P4 target as P4 switch.

\subsection{Local Controller}
We implement the local controller as Python 2.7 application.
\fig{implementation} depicts its parts.
We leverage the gRPC library \cite{grpc-python} to program the interfaces to the associated P4 switch and to the central controller.
We use the Scapy library \cite{scapy} to create and parse LLDP packets and the cryptography library \cite{cryptography} for applying AES-GCM to encrypt and decrypt LLDP packets.
For development and testing purposes, the local controller features a simple CLI.
It allows to write and read table entries and to display status changes on the P4 switch.

\subsection{Central Controller}
\begin{figure}[htp]
    \centering
    \includegraphics[width=0.85\linewidth]{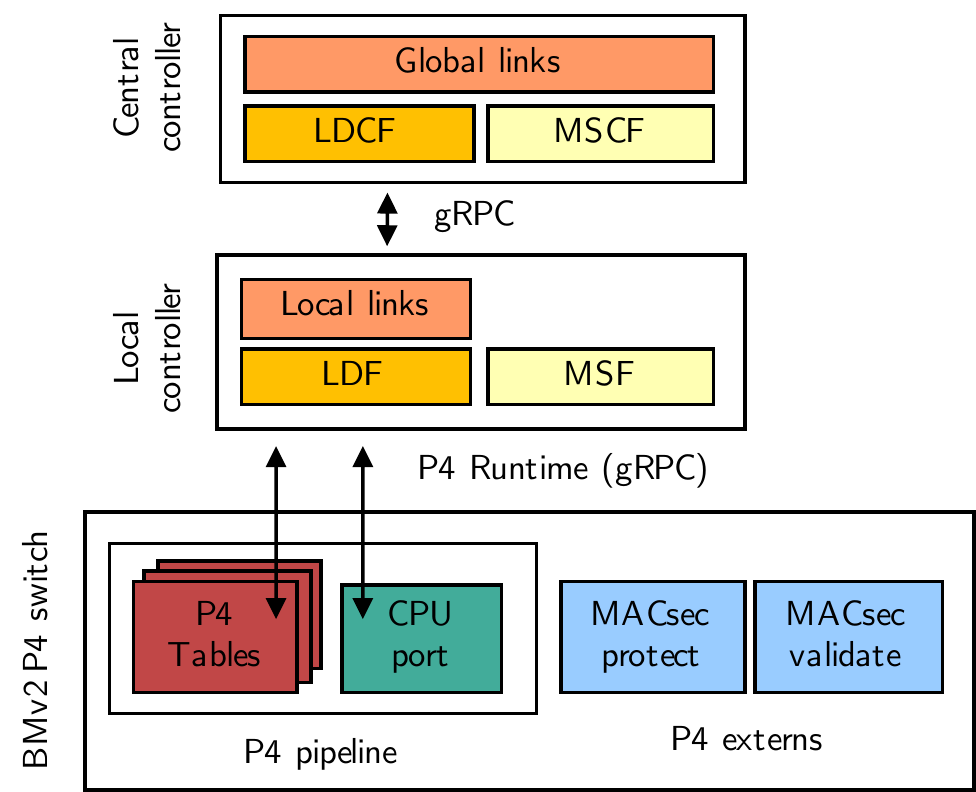}
    \caption{Structure of the prototypical implementation of MACsec. It consists of a BMv2 P4 software switch that implements the data plane functions of P4-MACsec and the two-tier control plane with the functions as introduced in \sect{concept}.}
    \label{fig:implementation}
\end{figure}

We implement the central controller similar to the local controller as Python 2.7 application.
\fig{implementation} depicts its parts.
It also leverages the gRPC library \cite{grpc-python} to build an gRPC interface to the local controller.
It also features a simple CLI for development and testing purposes that displays information about the current topology and active MACsec secure channels (SCs).
The control plane functions of P4-MACsec can also be integrated in ONOS \cite{onos} or other controller frameworks.
However, our objectives was a lightweight prototype using a slim and easy-to-understand controller implementation which directly leverages the P4 runtime and gRPC library for communication.
Thereby, we avoided dependencies on other controller frameworks which increase error space and implementation complexity.

  \section{Functional Validation}
\label{sec:validation}
We describe the experiment setup and validation experiments executed on the testbed from \sect{implementation-mininet}.

\subsection{Experiment I: Compliance to the MACsec Standard}
We first perform an experiment to examine the compliance of P4-MACsec to the IEEE 802.1AE standard.
Therefore, we set up a virtualized testbed that consists of a KVM virtual machine running Ubuntu Server in Version 18.04.1 LTS and our implementation of P4-MACsec on the BMv2 P4 software switch as described in \sect{implementation-mininet}.
The P4 switch connects another Ubuntu Server 18.04.1 LTS KVM/QEMU virtual machine that represents a network host behind a MACsec-enabled switch.
We configure a static MACsec connection between a P4 switch and a Linux host to check whether the MACsec implementation for BMv2 is compatible with the Linux implementation of MACsec.
On the Ubuntu server, we set up the static MACsec connection using the iproute2 tools.
For MACsec setup on the P4 switch, we use a simple Python script that adds the corresponding entries in the EG-SC, IG-SC, and SA tables of the P4 processing pipeline.
We successfuly validate that the Ubuntu server communicates with the P4 switch via MACsec in different communication scenarios, e.g., ICMP or streaming random data via TCP connections with netcat.
This does not validate a full compliance to all parts of the MACsec standard but demonstrates, that the P4 switch can communicate via MACsec with legacy devices.

\subsection{Experiment II: Complete P4-MACsec Scenario}
\begin{figure}[htp]
  \centering
  \includegraphics[width=.9\linewidth]{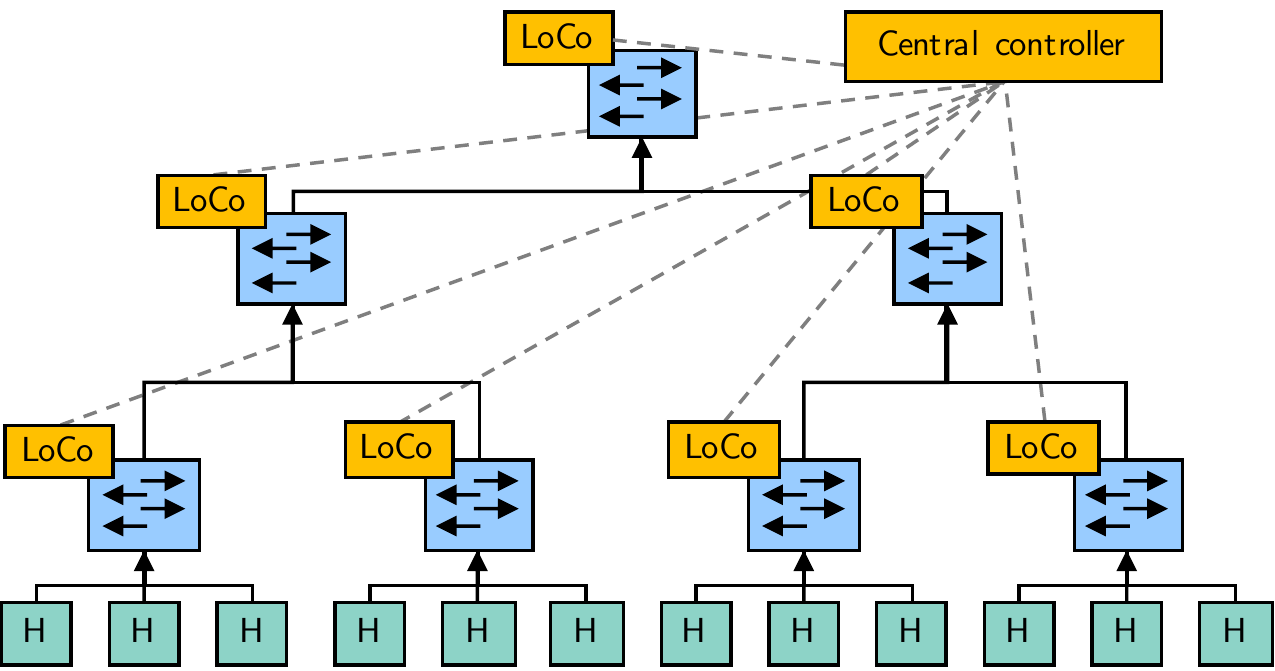}
  \caption{Virtual testbed environment that consists of network hosts, access switches, distribution switches, and a core switch. Each P4 switch is steered by a local controller (LoCo) that connects to the central controller.}
  \label{fig:testbed-environment}
  \vspace{-0.3cm}
\end{figure}

We now investigate the complete set of functionality of P4-MACsec.
Therefore, we create the topology depicted in \fig{testbed-environment}.
It follows the model of hierarchical network switches that consists of core, aggregation, and access switches.
A set of 12 network hosts is split into four groups, each attached to an access switch.
The four access switches are connected to two aggregation switches that are connected by a single core switch.
The testbed network is a single Layer 2 domain, i.e., network packets are forwarded based on their MAC address.
After starting the Mininet testbed, we verify the following aspects.
First, we examine that topology monitoring works correctly.
In initial link discovery, we verify that the detected topology matches the actual network topology.
Afterwards, we sporadically remove and re-add links between switches and supervise the process of link monitoring on the central controller via a CLI.
Second, we examine that automated deployment of MACsec and rekeying works correctly.
We investigate MACsec setup after changes in link monitoring through supervising the EG-SC, IG-SC, and SA tables on all P4-MACsec switches.
Last, we examine packet switching and correct setup of MACsec protection.
Therefore, we use ICMP and netcat to create network traffic between various pairs of network hosts in the experiment scenario.
We investigate packet traces on links between switches and verify that all packets are protected by MACsec.

  \section{Performance Evaluation}
\label{sec:evaluation}

We describe the evaluation setup and experiments for performance evaluation executed on the testbed from \sect{implementation-mininet}.

\subsection{Evaluation Setup}
\begin{figure}[htp]
    \centering
    \includegraphics[width=0.98\linewidth]{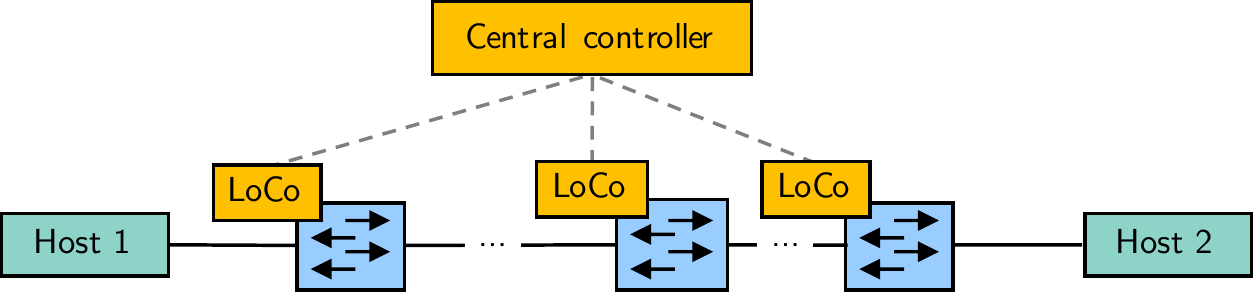}
    \caption{Evaluation testbed that consists of two network hosts that are attached to two P4 switches. In between, we vary from $0$ to $6$ additional P4 switches to form variable-length P4 switch chains for the evaluation experiments.}
    \label{fig:evaluation-setup}
\end{figure}

\fig{evaluation-setup} depicts the evaluation setup.
It consists of two network hosts that are attached to two P4 switches with $0$ to $6$ P4 switches in between.
We perform performance evaluation experiments to investigate the throughput and round-trip time (RTT).
We vary the number of P4 switches between the two network hosts and measure throughput and RTT for $1$ to $8$ hops.
For each evaluation experiment, we consider three scenarios.
In the first scenario, MACsec is disabled, i.e., the P4 switches between Host 1 and Host 2 only perform MAC address learning and L2 forwarding.
In the second scenario, we enable MACsec so that all packets between Host 1 and Host 2 are protected with AES-GCM implemented as P4 Extern.
In the third scenario, we enable MACsec but skip AES-GCM encryption and decryption in the P4 extern so that only cleartext payloads are sent within the MACsec packets.
This third scenario is not part of the MACsec standard but a modification for the performance evaluation that allows us to measure the impact of AES-GCM in comparison to the impact of exchanging network packets with an P4 extern.

\subsection{TCP Throughput}
\begin{figure}[htp]
  \centering
  \includegraphics[width=0.85\linewidth]{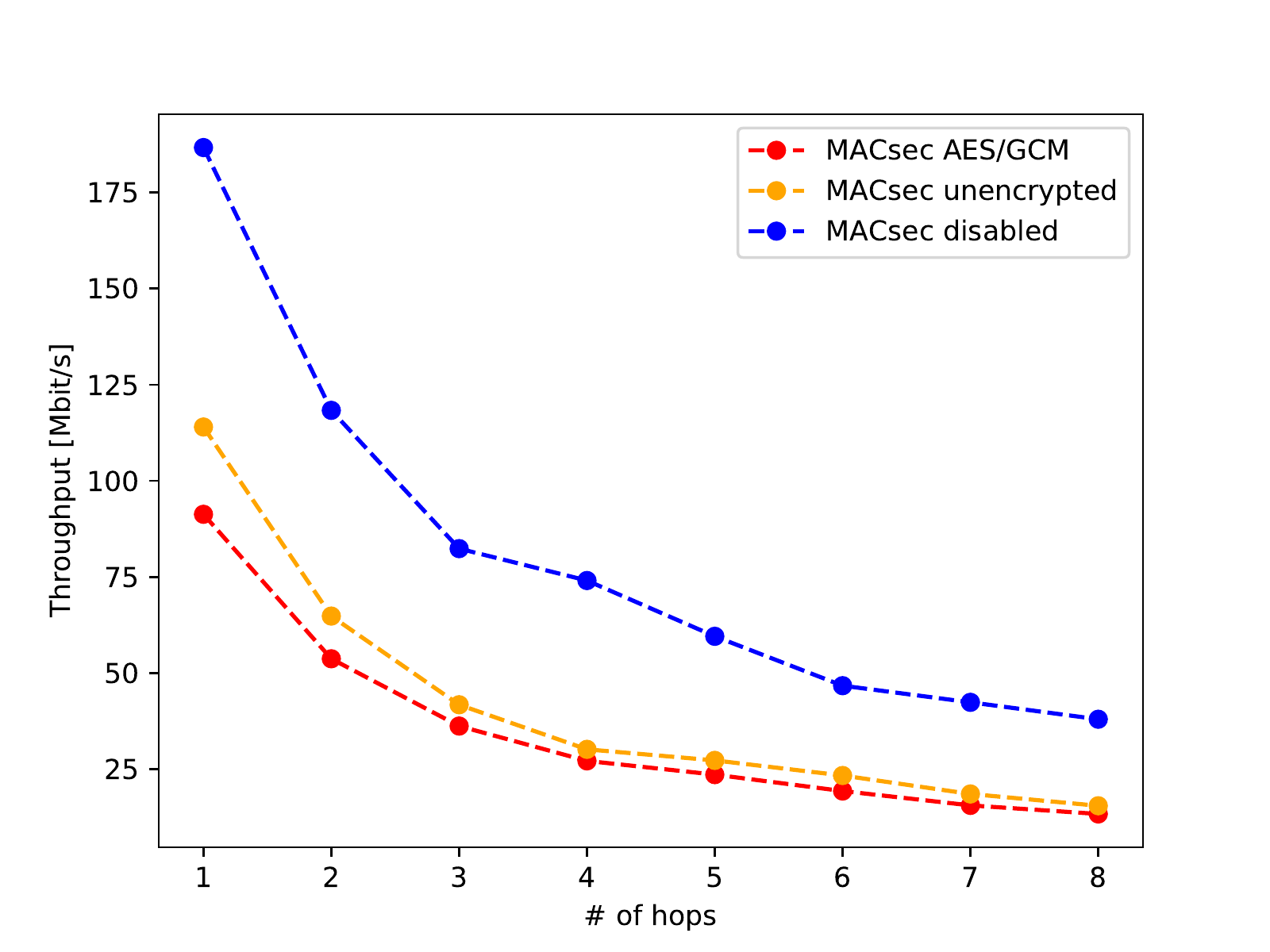}
  \caption{TCP throughput evaluation with $1$ to $8$ hops represented by P4 switches between two network hosts with iperf3. We consider three scenarios: disabled MACsec, enabled MACsec with encryption and decryption using AES-GCM, and enabled MACsec without encryption and decryption.}
  \label{fig:throughput}
\end{figure}

We first investigate on the TCP throughput in P4-MACsec.
Therefore, we measure TCP transmissions between Host 1 and Host 2 with iperf3 \cite{iperf3}.
Host 1 runs an iperf server, Host 2 runs an iperf client.
We perform three runs, each with a duration of $\textrm{30\ seconds}$.
\fig{throughput} depicts the results calculated as average over the three runs.
As expected, the throughput for all scenarios decreases with the number of P4 switches that run on the testbed.
Enabled MACsec processing causes a large degradation in throughput.
However, applying or omitting AES-GCM in the P4 extern does not cause large differences in throughput.
The large decrease in throughput is not the result of encryption or decryption with AES-GCM but an effect of the interaction between the P4 pipeline and the MACsec protect and MACsec validate P4 externs in BMv2.

\subsection{Round-Trip Time (RTT)}
\begin{figure}[htp]
  \centering
  \includegraphics[width=0.85\linewidth]{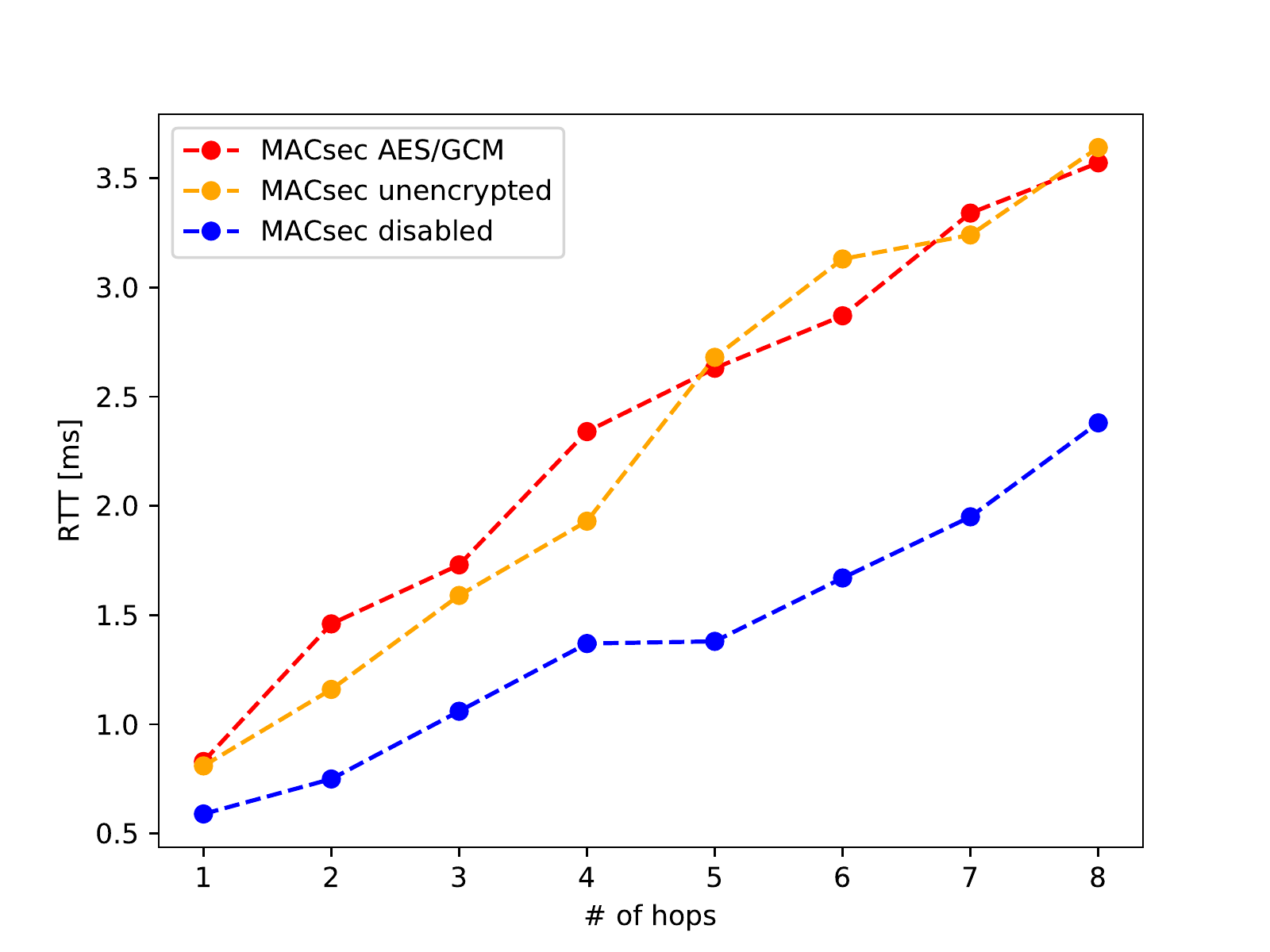}
  \caption{Round-trip time (RTT) evaluation with $1$ to $8$ hops represented by the P4 switches between two network hosts with ping. We consider three scenarios: disabled MACsec, enabled MACsec with encryption and decryption using AES-GCM, and enabled MACsec without encryption and decryption.}
  \label{fig:rtt}
\end{figure}

In the second experiment, we investigate the round-trip time (RTT) between the two network hosts that are connected by $1$ to $8$ P4 switches in between.
We use the ping \cite{ping} tool on Host 1 to send $1000$ consecutive ICMP echo requests to Host 2.
We set an idle period of $0.01$ seconds between two ICMP packets and perform three runs of the experiment.
\fig{rtt} depicts the RTTs calculated as average over the three runs.
The evaluation results are similar to those of the experiment on TCP throughput.
Enabled MACsec causes an increase of the RTT.
Again, applying or omitting AES-GCM in the P4 extern does not cause large differences in the RTT.
As in the experiment on TCP throughput, the interaction between the P4 pipeline and the MACsec protect and MACsec validate P4 externs in BMv2 seems to cause the negative effects on the RTT.

\subsection{MACsec Setup Time}
As described in \sect{concept}, the two-tier control plane automatically configures and enables MACsec on all assigned P4 switches.
As described in \sect{macsec-mka}, MACsec deployments require configuration effort whether static key setup on all switches or MKA is applied.
With P4-MACsec, that configuration effort to deploy MACsec completely disappears.
The process of link discovery, MACsec setup, and the periodically reiterations are performed within nearly not measureable time intervals.
The two-tier control plane performs all actions sequentially, i.e., delays might turn up with a very large number of controlled P4 switches.
  \section{Implementation on NetFPGA SUME}
\label{sec:implementation-hardware}

In the following, we briefly describe the NetFPGA SUME \cite{netfpga-website} platform and outline our experiences in implementing P4-MACsec for that platform.

\subsection{NetFPGA SUME Platform}
The NetFPGA SUME board is a platform for rapid prototyping of network applications with bandwidths up to 10 Gb/s.
It features a Virtex-7 690T field-programmable gate array (FPGA), four SFP+ network transceivers, and an PCI Express interface to the host system \cite{netfpga-ieee}.
The P4-NetFPGA project \cite{p4-netfpga} transforms the NetFPGA SUME board into a hardware P4 switch.
P4 programs are transformed into SDNet descriptions by the P4-SDNet compiler that creates HDL modules that run as part of the reference architecture of the NetFPGA SUME board.

\subsection{Implementation of P4-MACsec}
We modify the P4 processing pipeline of our software prototype to cope with limitations of the architecture, e.g., a missing lookahead function in packet parsing or the limitation to a single instead of multiple control blocks in the P4 processing pipeline.
We implemented AES-GCM based on a publicly available Verilog module from OpenCores \cite{aesgcm-opencores}.
However, we were not able to create a fully working P4-MACsec switch due to two severe limitations.
First, the NetFPGA SUME platform does not provide functions to parse or access variable-length payloads of network packets.
Therefore, payloads of network packets need to be parsed as headers, which limits the implementation to fixed-length packets.
Last, exchange of packet data between the P4 processing pipeline and the P4 external function is limited.
Currently, data that is transferred from the P4 processing pipeline to a P4 external function needs to be transmitted within one clock cycle of the FPGA.
Due to timing limitations, it is only possible to transmit very small amounts of data.
The developers from the P4-NetFPGA project confirmed that the current version does not allow to process complete network packets within P4 externs.
We were able to increase the amount of data to be exchangeable by reducing the base clock frequency of the NetFPGA to 128 bytes.
However, this is still far away from real-world applicability.
A packet streaming function through P4 external functions was announced, but is not available so far.
Summing up, both limitations did not allow us to build a prototype that is suitable for real-world scenarios with variable-length packets exceeding a total length of 128 bytes.
  \section{Conclusion}
\label{sec:conclusion}

In this work we proposed P4-MACsec, a concept to automatically protect links between switches with MACsec in P4-SDN.
Our concept features a P4 data plane implementation for MACsec including encryption and decryption using AES-GCM.
P4 switches are steered by a novel two-tier control plane that consists of local controllers running on all P4 switches that connect to a central controller.
We presented a novel mechanism for link discovery using encrypted LLDP packets and automated deployment of MACsec link protection.
We presented the architecture of P4-MACsec and demonstrated its feasability in a prototypical implementation for the BMv2 P4 software switch.
We used that prototype to experimentally validate P4-MACsec in a virtualized testbed built with Mininet and performed evaluation experiments.
We discussed experiences with implementing P4-MACsec for the NetFPGA SUME platform.
We discovered that both P4 targets have major constraints regarding the implementation of P4-MACsec.
Using the P4 externs for MACsec on the NetFPGA SUME platform was not feasible at all.
Only fixed-lenght packets that do not exceed a total size of 128 bytes can be exchanged.
Using the P4 externs for MACsec on the BMv2 P4 target results in large degradations in TCP througput and latency in RTTs.
However, use-cases that aim at securing the network by applying authentication, encryption, and integrity checks are important.
P4-MACsec completely eliminates previous configuration efforts for MACsec as network security mechanism.
Similar approaches for traffic protection on different layers, e.g., L3 or L4 VPN, are imaginable.
Therefore, P4 switches should offer native functional blocks for encryption and decryption and overhead-free interfaces to P4 externs.
  \section*{Acknowledgment}
The authors thank Joshua Harmann for fruitful discussions and programming contributions.

  \bibliographystyle{IEEEtran}
  \bibliography{paper}
  
  \balance
  
  \end{document}